 \documentclass[12pt]{article}
 \begin{document}
 \title{Mathematical genesis ?\\Big Bang and Cantor set}
 \author{H. Tasso \\Max-Planck-Institut f\"{u}r Plasmaphysik\\Euratom
 Association\\ 85748 Garching bei M\"{u}nchen, 
 Germany}
 \maketitle

 \begin{abstract} A mathematical framework is proposed for the "big bang". It 
starts with some Cantor set and assumes the existence of a transformation from 
that set to the 
continuum set used in conventional physical theories like general relativity and 
quantum mechanics. 
\end{abstract}
 
Let us assume that the big bang started from a Cantor set (see e.g. \cite{wik}) 
with the power of continuum and an arbitrary cardinality higher than the one of 
countable sets (see e.g. \cite{mcg}). 

Since a Cantor set has zero measure, it can be considered as physically empty. 
However, if the inverse construction out of a Cantor set to some sets of the 
reals exists, for example 
by appropriate "rearrangement" of its elements, then a nonzero measure can be 
produced and, possibly, the continuum of the physical world.

This is not unlikely because the simplest Cantor set \cite{wik} has the same 
power as the interval [0,1], though, originally, it has been constructed by 
annihilating the measure of that interval. Moreover it is known that a 
continuous surjection from the Cantor set to the interval [0,1] exists 
\cite{wik}. However, in general, such transformations can imply very 
sophisticated tools for an infinite set (see e.g. \cite{had}).

A nonvanishing measure is essential to introduce spacetime and Hilbert space 
in order to formulate general relativity and quantum 
mechanics. Also the theory of "superstrings" could be the result of such a 
procedure, but alternative theories could be described in the same frame as 
well. 

The physical questions arising within the first millisecond \cite{ree} of the 
big bang are then related 
to the nature of the nontrivial "rearrangement" mentioned above and to the 
choice of cardinality of the Cantor set. This would be a framework permitting 
somehow to describe "everything out of almost nothing" to compare with ideas  
suggested in Ref. \cite{whe}. 

On the other hand, the mere existence of "rearrangements" does not explain, for 
example, 
the inflation in size of the early universe needed in order to match the 
backward calculations (see e.g. \cite{ree}) from the present universe  to its 
size at about a millisecond after the big bang. Such an inflation corresponds in 
our scheme to an "explosion" of measure, which is similar to a "phase change" or 
an instability triggered by a specific parameter like "temperature" in 
statistical mechanics or thermodynamics. A free parameter here could be the 
cardinality of the Cantor set whose special choice could activate the "phase 
change" from the Cantor set to the desired four-dimensional early universe. 
 
Finally, though the physics within the first millisecond of the big bang is not 
known, it is expected that it should contain the embryo of the physics of the 
present universe. The solution to this problem remains a long term project.

\begin{center}
{\bf ACKNOWLEDGEMENTS}
\end{center}

The author would like to thank Labib Haddad and Rudolf Gorenflo for clarifying 
discussions on set theory.

\end{document}